\documentclass[english,aps,twocolumn,prl,superscriptaddress]{revtex4}
\usepackage{exscale}
\usepackage{graphicx}
\usepackage{amsmath}
\usepackage{latexsym}
\usepackage{amsfonts}
\usepackage{amssymb}
\usepackage{amscd}
\usepackage{bbm}
\usepackage{dcolumn}      
\usepackage{bm}           
\usepackage{pstricks,pst-grad,fancybox,graphics}
\usepackage{enumerate}
\usepackage{epstopdf}
\usepackage[T1]{fontenc}
\usepackage[latin9]{inputenc}

\newcommand{\ket}[1]{\ensuremath{|#1\rangle}}

\newcommand{\BE}{\begin{equation}}
\newcommand{\EE}{\end{equation}}
\newcommand{\be}{\begin{equation}}
\newcommand{\ee}{\end{equation}}
\newcommand{\bea}{\begin{eqnarray}}
\newcommand{\eea}{\end{eqnarray}}
\newcommand{\bean}{\begin{eqnarray*}}
\newcommand{\eean}{\end{eqnarray*}}
\newcommand{\kommentar}[1]{}

\newcommand{\mean}[1]{\ensuremath{\langle #1 \rangle}}

\newcommand{\bc}{\begin{center}}
\newcommand{\ec}{\end{center}}

\newcommand{\ud}{\mathrm{d}}

\usepackage[normalem]{ulem}



\usepackage{babel}

\begin{document}

\title{Useful multiparticle entanglement and sub shot-noise sensitivity\\ in experimental phase estimation}

\author{Roland~Krischek} 
\affiliation{Fakult\"at f\"ur Physik, Ludwig-Maximilians Universit\"at M\"unchen, D-80799 M\"unchen, Germany}
\affiliation{Max-Planck Institut f\"ur Quantenoptik, Hans-Kopfermann Str. 1, D-85748 Garching, Germany}
\author{Christian~Schwemmer} 
\affiliation{Fakult\"at f\"ur Physik, Ludwig-Maximilians Universit\"at M\"unchen, D-80799 M\"unchen, Germany}
\affiliation{Max-Planck Institut f\"ur Quantenoptik, Hans-Kopfermann Str. 1, D-85748 Garching, Germany}
\author{Witlef~Wieczorek}
\altaffiliation{Present address: Vienna Center for  
Quantum Science and Technology, Faculty of Physics, University of Vienna,  
Boltzmanngasse 5, A-1090 Vienna, Austria}
\affiliation{Fakult\"at f\"ur Physik, Ludwig-Maximilians Universit\"at M\"unchen, D-80799 M\"unchen, Germany}
\affiliation{Max-Planck Institut f\"ur Quantenoptik, Hans-Kopfermann Str. 1, D-85748 Garching, Germany}
\author{Harald~Weinfurter} 
\affiliation{Fakult\"at f\"ur Physik, Ludwig-Maximilians Universit\"at M\"unchen, D-80799 M\"unchen, Germany}
\affiliation{Max-Planck Institut f\"ur Quantenoptik, Hans-Kopfermann Str. 1, D-85748 Garching, Germany}
\author{Philipp~Hyllus}
\altaffiliation{Present address: 
Dep. of Theor. Phys., The Univ. of the Basque Country, P.O. Box 644, E-48080 Bilbao, Spain}
\affiliation{INO-CNR BEC Center and Dipartimento di Fisica, Universit{\`a} di Trento, I-38123 Povo, Italy}
\author{Luca~Pezz{\'e}} 
\affiliation{Laboratoire Charles Fabry de l'Institut d'Optique, CNRS and Universit\'e Paris-Sud, 
F-91127 Palaiseau cedex, France}
\author{Augusto~Smerzi}
\affiliation{INO-CNR BEC Center and Dipartimento di Fisica, Universit{\`a} di Trento, I-38123 Povo, Italy}

\begin{abstract}
We experimentally demonstrate a general criterion to identify entangled 
states useful for the estimation of an unknown phase shift with a sensitivity 
higher than the shot-noise limit.
We show how to exploit this entanglement on the examples of
a maximum likelihood as well as of a Bayesian phase estimation protocol.
Using an entangled four-photon state we achieve a phase sensitivity 
clearly beyond the shot-noise limit. Our detailed comparison 
of methods and quantum states for entanglement enhanced metrology reveals the 
connection between multiparticle entanglement and sub shot-noise uncertainty, 
both in a frequentist and in a Bayesian phase estimation setting.
\end{abstract}



\maketitle

\noindent

The field of quantum enhanced metrology is attracting increasing interest \cite{GiovannettiNatPhot11}
and impressive experimental progress has been achieved
with photons \cite{RarityPRL90, MitchellNat04,KacprowiczNP2010, NagataSci07,  Xiang_2010},
cold/thermal atoms \cite{AppelPNAS2009},
ions \cite{LeibfriedSci04} and Bose-Einstein
condensates \cite{GrossNat10, RiedelNat10}. 
Several experiments have demonstrated phase super resolution \cite{MitchellNat04, LeibfriedSci04}, 
which, if observed with a high visibility of the interference fringes, allows
to utilize the state for quantum enhanced metrology \cite{ReschPRL07,NagataSci07}.
So far, only few experiments have implemented a 
full phase estimation protocol beating the shot-noise 
limit $\Delta\theta=1/\sqrt{N}$ with $N>2$, where $N$ is the number of particles
\cite{LeibfriedSci04, AppelPNAS2009, GrossNat10}.
Recently, it has been theoretically shown that sub 
shot-noise (SSN) phase sensitivity requires the presence 
of (multi-)particle entanglement \cite{PezzePRL09,HyllusArXiv10b}. 
In this letter, we experimentally demonstrate 
this connection. For an entangled state and a separable state with 
$N=4$ addressable photons, we measure the quantum Fisher information (QFI) 
$F_Q$ \cite{Helstrom67}, which quantifies the amount of entanglement 
of the state useful for SSN interferometry \cite{PezzePRL09}.
We then show how this entanglement can indeed be exploited by 
implementing a Maximum Likelihood (ML) and a Bayesian phase estimation
protocol, both clearly yielding SSN phase uncertainty.



The usefulness of an experimental state  
can be quantified by the quantum Fisher
information (QFI) $F_Q[\rho,\hat J]$ \cite{Helstrom67}.
A probe state $\rho$ of $N$ qubits is entangled {\em and}
allows for SSN phase estimation if the condition
\be \label{eq:FQN}
	F_Q[\rho,\hat J] > N
\ee
is fulfilled \cite{PezzePRL09}. 
Here $\hat J=\frac{1}{2}\sum_{i=1}^N \hat\sigma_{\vec n_i}^{(i)}$
is the linear generator of the phase shift, and 
$\hat\sigma_{\vec n_i}^{(i)}= \vec n_i \cdot \hat \sigma$ 
is a Pauli matrix rotating the qubit $i$ along the 
arbitrary direction ${\vec n}_i$.
The maximal $F_Q$ further depends
on the hierarchical entanglement structure of the probe state and
genuine multiparticle entanglement is needed to reach 
the Heisenberg limit \cite{HyllusArXiv10b,SI}, 
the ultimate sensitivity allowed by quantum mechanics.
With $N=4$ qubits, 2-particle entangled states have $F_Q\le 8$, 
while for 3-particle entangled states $F_Q\le 10$ \cite{HyllusArXiv10b,nota_k-ent}.
The ultimate limit is $F_Q \le N^2=16$ which is saturated by the so-called Greenberger-Horne-Zeilinger (GHZ) state
\cite{GHZ,GiovannettiPRL06,PezzePRL09}.

A state fulfilling Eq.~(\ref{eq:FQN}) allows for SSN phase uncertainty 
due to the Cramer-Rao theorem, which limits
the standard deviation $\Delta \theta$ of 
unbiased phase estimation as \cite{Cramer_book, Helstrom67, note_resources}
\be
	\label{eq:CR}
	\Delta \theta \ge \frac{1}{\sqrt{m F_{\hat \mu}\big[\theta_0,\rho,\hat J\big]}}
	\ge \frac{1}{\sqrt{m F_Q\big[\rho,\hat J\big]}}.
\ee
The first inequality defines the Cramer-Rao lower bound (CRLB).
Here $\theta_0$ is the true value of the phase shift,
$m$ is the number of repeated independent measurements, and 
\be \label{eq:Fisher}
	F_{\hat{\mu}}\big[ \theta_0,\rho,\hat J \big]=
	\sum_{\mu} \frac{1}{P(\mu|\theta_0)} \Big(\frac{\ud P(\mu|\theta)}{\ud \theta}\Big|_{\theta_0} \Big)^2
	\leq F_{Q}\big[\rho,\hat J \big].
\ee
The Fisher information $F_{\hat{\mu}}\big[ \theta_0,\rho,\hat J \big]$ depends 
on the conditional probabilities $P(\mu|\theta_0)$
to obtain the result $\mu$ in a measurement 
when the true phase shift is equal to $\theta_0$.
It is bounded by the QFI \cite{PezzePRL09,Helstrom67},
the equality being saturated for an optimal measurement $\hat\mu_{\rm opt}$.
From Eqs~(\ref{eq:FQN}) and (\ref{eq:CR}) and from the bounds for multi-particle entanglement, 
we can infer that, if the experimentally obtained $F_Q$ of a $N$-qubit state exceeds the value for $k$-particle 
entanglement, one can achieve a phase sensitivity better than that achievable with any $(k-1)$-particle entangled state of any $N$ qubits \cite{nota_k-ent}.


\begin{figure}[t!]
\begin{center}
  \includegraphics[clip,scale=0.9]{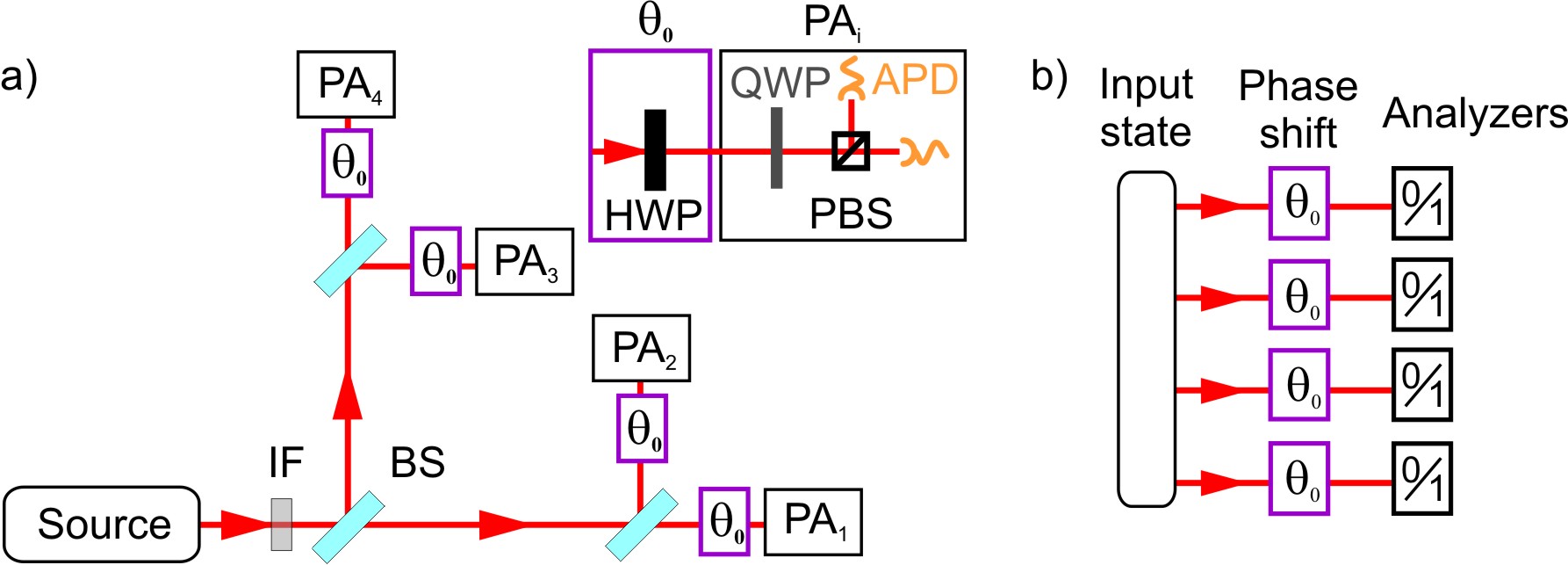}
\end{center}
\caption{a) Experimental setup. 
The source uses pulsed parametric down conversion with a type II 
cut $\beta$-Barium-Borate crystal ($\lambda_{pump}=390$nm) \cite{SI}.
After passing an interference filter (IF), 
the photons are symmetrically distributed into 4 spatial modes by using 3
non-polarizing beam splitters (BS).
The Dicke state $\ket{D_4^{(2)}}$ is observed 
if one photon is detected in each of the four output 
arms \cite{WieczorekPRL09}.
The separable state $\ket{\psi_{\rm sep}}$ is created 
by inserting a $\ket{+}$ polarizer before the first BS.
Each polarization qubit is addressed individually and
rotated by $\exp[-i \hat{\sigma}_y \theta/2]$ (violet box)
by a halfwave-plate (HWP).
Each polarization analyzer (PA) is composed of a 
quarterwave-plate (QWP), a polarizing beam-splitter (PBS) and an
avalanche photo-diode (APD).
b) Schematic of our interferometric setup.}  \label{fig:setup}
\end{figure}

For the experimental demonstration,
we use the symmetric four-photon entangled Dicke 
state \cite{KieselPRL07, notaFock} 
$\ket{D_4^{(2)}}=\big(\ket{HHVV}+\ket{HVHV}+\ket{HVVH}+\ket{VHHV}+\ket{VHVH}+\ket{VVHH}\big)/\sqrt{6}$
and the separable state $\ket{\psi_{\rm sep}} = \ket{++++}$ observed from multiphoton parametric down conversion \cite{WieczorekPRL09} [Fig.~\ref{fig:setup} a)].
Here $\ket{HHVV} \equiv \ket{H}_1\otimes\ket{H}_2\otimes\ket{V}_3\otimes\ket{V}_4$,
$\ket{H}_i$ ($\ket{V}_i$) refer to the horizontal (vertical) polarization of a
photon in the spatial mode $i$, and $\ket{\pm} \equiv\frac{1}{\sqrt{2}}(\ket{H}\pm\ket{V})$.
From the measured density matrices ($\rho^{\rm exp}_D$ and $\rho^{\rm exp}_{\rm sep}$ \cite{SI}) 
we deduce a fidelity of $0.8872\pm 0.0055$ for $\ket{D_4^{(2)}}$ and $0.9859 \pm 0.0062$ for $\ket{\psi_{\rm sep}}$ (errors 
deduced with Poissonian count statistics) 
and also the QFI determining 
the suitability of the experimentally observed states for phase estimation.
For the ideal Dicke state $\ket{D_4^{(2)}}$, the QFI reaches its maximum value, 
$F_Q[\ket{D_N^{(N/2)}},\hat J]=N(N+2)/2=12$, when
$\hat \sigma_{\vec n_i}=\hat\sigma_y$ for all $i$ ($\hat J \equiv \hat J_y$) \cite{HyllusPRA10}. 
In the experiment, this choice leads to $F_Q[\rho^{\rm exp}_D,\hat J_y]=9.999\pm 0.095$, 
at the maximal value achievable with 3-particle entanglement. 
An optimization over the local directions ${\vec n_i}$~\cite{HyllusPRA10}, 
leads to the slightly higher value $F_Q^{\rm opt}[\rho^{\rm exp}_D,\hat J^{\rm opt}]=10.326\pm 0.093$, detecting useful 4-particle entanglement
with $3.5$ standard deviations.
Sure enough, using a witness operator it is possible
to prove 4-particle entanglement in a simpler way 
\cite{KieselPRL07,SI}. With only a subset of the tomographic
data we obtain a witness expectation value of
$-0.2205\pm 0.0055$, proving 4-particle entanglement
with a significance of 40 standard deviations \cite{SI}.
However, witness operators merely recognize entanglement, 
whereas our criterion directly indicates
the state's applicability for a quantum task.
The separable state $\ket{\psi_{\rm sep}}$ ideally allows for 
sensitivity at the shot-noise limit,
$F_Q[\ket{\psi_{\rm sep}},\hat J_y]=N=4$.
The experimental density matrix leads to $F_Q[\rho^{\rm exp}_{\rm sep},\hat J_y]=3.894\pm 0.023$, 
a value close to the expected separable limit
(the optimized value being
$F_Q^{\rm opt}[\rho^{\rm exp}_D,\hat J^{\rm opt}]=4.014\pm 0.025$).

In order to demonstrate that the precision close to the one
predicted by $F_Q$ can indeed be achieved in practise,
we experimentally implement a phase estimation analysis with 
the input states $\rho^{\rm exp}_D$ and $\rho^{\rm exp}_{\rm sep}$.
Our interferometric protocol transforms the probe state
by $U(\theta_0)=\exp[-i\sum_{k=1}^4 \hat\sigma_{\vec{n}_i}^{(k)} \theta_0/2]$ using the 
halfwave-plates and phase shifts depicted in Figs~\ref{fig:setup} a) and
b).
The unknown value of the phase shift $\theta_0$ is inferred 
from the difference in the number of particles, $2\mu=N_H-N_V$ ($\mu=-2,-1,0,1,2$), 
in the states $V$ and $H$.
For the ideal states
and the rotation directions $\vec{n}_i=y$, 
this measurement is optimal, and hence $F_{\hat \mu}=F_Q$.
Experimentally, the optimized direction and measurement 
can be different because of noise and misalignment.
However, for the observed states 
the expected improvement would be rather small.

\begin{figure*}[t!]    	
\begin{center}
  \includegraphics[clip,width=0.9\linewidth]{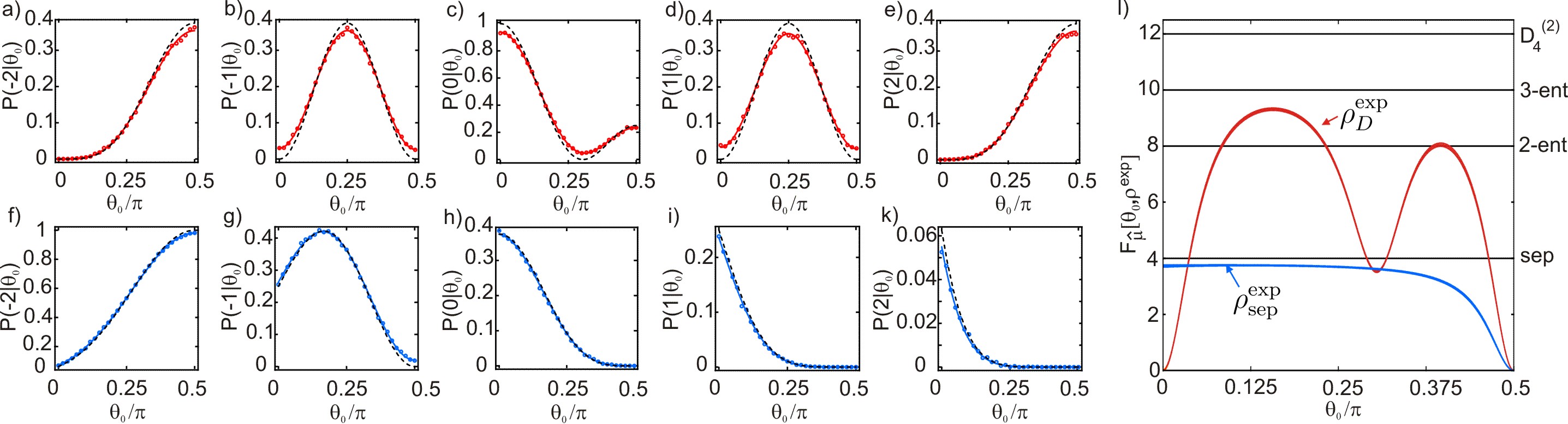}
\end{center}
 \caption{ 
 Calibration curves and Fisher information. 
 The small panels show the conditional probabilities 
 $P_{\rm exp}(\mu|\theta_0)$ for the state $\rho_D^{\rm exp}$ [red curve, upper row a)-e)]
 and for $\rho_{\rm sep}^{\rm exp}$ [blue curve, lower row f)-k)].
 Dashed black lines are the ideal probabilities $P(\mu|\theta)$, 
 dots are experimental results. 
 The red and blue curves are fits obtained by assuming that 
 the main source of errors are misalignments in the 
 polarization optics \cite{SI}.
 The measurements are performed for 31 values of $\theta_0$ by collecting approximately 7000 
 events for each phase value.
 Panel l) shows the Fisher information [Eq.~(\ref{eq:Fisher})], 
 obtained from the fits $P_{\rm exp}(\mu|\theta_0)$.
 The line widths correspond to the error intervals 
 $F_{\hat \mu}[\theta_0,\rho^{\rm exp}]\pm\Delta F_{\hat \mu}$
 with $\Delta F_\mu[\rho_D^{\rm exp}]\le 0.08$ and 
 $\Delta F_{\hat\mu}[\rho_{\rm sep}^{\rm exp}]\le 0.04$ \cite{SI}.
 Horizontal lines indicate limits for separable states
 (``sep'', equal to $F_{\hat\mu}[\ket{\psi_{\rm sep}}]$), for  
 2- and 3-particle entangled states, and for the ideal Dicke state 
 $F_{\hat \mu}[\ket{\mathrm{D}_4^{(2)}}]$. Theoretically, $F_{\hat\mu}=F_Q$ holds for the ideal 
 input states, phase operations and output measurements. 
Experimentally, we observe $F_{\hat \mu}[\theta_0,\rho^{\rm exp}] < F_Q[\rho^{\rm exp}]$ due to technical
noise.
 In particular, $F_{\hat\mu}[\theta_0,\rho^{\rm exp}_D]$ is strongly reduced for values of the phase shift where some of the ideal conditional probability densities $P(\mu|\theta_0)$ of panels a)-e) go to $0$ ($\theta_0=0,0.3\pi,0.5\pi$). 
 For reduced visibilities (when $P_{\rm exp}(\mu|\theta_0)>0$ 
 while ideally $P(\mu|\theta_0)=0$),
 the contribution to the Fisher information is reduced since in
 these points also the derivatives of $P_{\rm exp}(\mu|\theta_0)$ vanish, 
 cf. Eq.~(\ref{eq:Fisher}). 
} \label{fig:Pfit}
\end{figure*}

The relation between the phase shift and the 
possible results of a measurement is provided by the 
conditional probabilities $P(\mu|\theta_0)$.
These are measured experimentally and compared with the theoretical ones
for both the separable and the entangled state, as shown 
in Fig.~\ref{fig:Pfit} a)-k).
A fit to the measured conditional probabilities provides $P_{\rm exp}(\mu|\theta)$, 
which are used to calculate the Fisher information 
according to Eq.~(\ref{eq:Fisher}) [see Fig.~\ref{fig:Pfit} l)].
As expected, our experimental apparatus can surpass the shot noise limit 
for a broad range of phase values (where $F_{\hat\mu}^{\rm exp} > 4$), 
and can even exploit useful three particle entanglement (where $F_{\hat\mu}^{\rm exp} > 8$).

The phase shift $\theta_0$ is inferred from the 
results, $\mu_1,\mu_2,...,\mu_m$, of $m$ independent 
repetitions of the interferometric protocol.
We will refer to such a collection of 
measurements as a single $m$-experiment.
In the experiment, we set the phase shift to 9 known values $\theta_0$. 
For each $\theta_0$, 12000 results $\mu_i$ are independently measured and grouped into vectors of length $m$ to
perform the phase estimation for different values 
of $m$ $(=1,10,100)$. Using this data, we implement a ML and a 
Bayesian phase estimation protocol. 
While both have been recently used in
literature for phase estimation \cite{KacprowiczNP2010, PezzePRL07},
here they are compared in detail and applied for the first time to demonstrate SSN phase uncertainty with more than two particles. To display the quantum enhancement and to compare the methods we use the rescaled uncertainty $\Delta_{\rm res}$ defined below.

In the ML protocol, the estimator $\theta_{\mathrm{est}}$ of the unknown phase shift is 
determined as the value maximizing the likelihood function 
${\cal L}(\theta)=\prod_{i=1}^m P_{\rm exp}(\mu_i|\theta)$ \cite{Cramer_book}.
For different $m$-experiments it fluctuates
with standard deviation $\Delta \theta_{\mathrm{est}}$, which has 
to be calculated by repeating
a large number 
of single $m$-experiments.
For large $m$, the distribution of $\theta_{\mathrm{est}}$ approaches 
a Gaussian centered on $\theta_0$ and of width $\Delta \theta_{\mathrm{est}}$  
saturating the CRLB,
Eq.~(\ref{eq:CR}) \cite{Cramer_book}. 

\begin{figure}[t!]
\begin{center}
  \includegraphics[clip,width=0.9\linewidth]{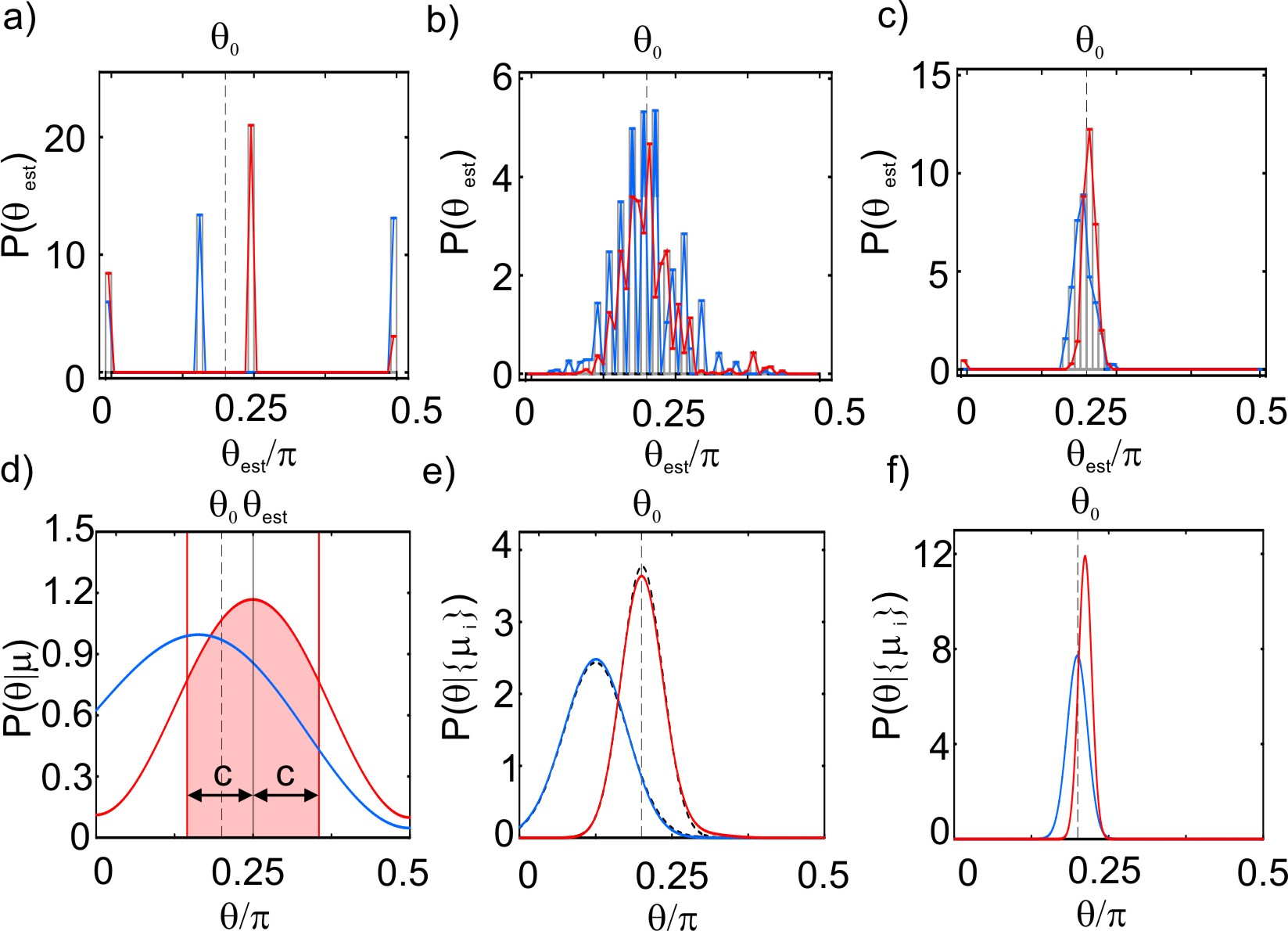}
\end{center}
 \caption{Comparison of the ML method to the Bayesian approach
 for the estimation of a phase shift $\theta_0=0.2\pi$ (vertical dashed black line). 
 Upper row: histograms (normalized to one) of the estimators $\theta_{\mathrm{est}}$
 obtained for large number of repetitions of $m$-experiments: 
 a) $m=1$, b) $m=10$, and c) $m=100$.
 Red (blue) solid lines show the results of the state $\rho_D^{\rm exp}$ ($\rho_{\rm sep}^{\rm exp}$). 
 Lower row: exemplary Bayesian probability densities 
 $P(\theta|\{\mu_i\}_{i=1}^m)$
 of {\em single} $m$-experiments
 for d) m=1, e) m=10, and f) m=100 
 for the state $\rho_D^{\rm exp}$ (solid red lines) and 
 $\rho^{\rm exp}_{\rm sep}$ (solid blue lines). In panel e)
 the dashed black lines are Gaussians 
 of width $1/\sqrt{m\,F_{\hat\mu}}$ plotted to illustrate 
 that the densities rapidly approach a Gaussian shape. 
 For $m=1$, we plot $P(\theta|\mu=1)$ for $\rho_D^{\rm exp}$
 and $P(\theta|\mu=-1)$ for $\rho^{\rm exp}_{\rm sep}$.
 The shaded region indicates the confidence interval 
 $[\theta_{\mathrm{est}}-C,\theta_{\mathrm{est}}+C]$
 around the maximum of the distribution.}
 \label{fig:histograms}
\end{figure}

\begin{figure}[b!]
\begin{center}
  \includegraphics[clip,width=0.9\linewidth]{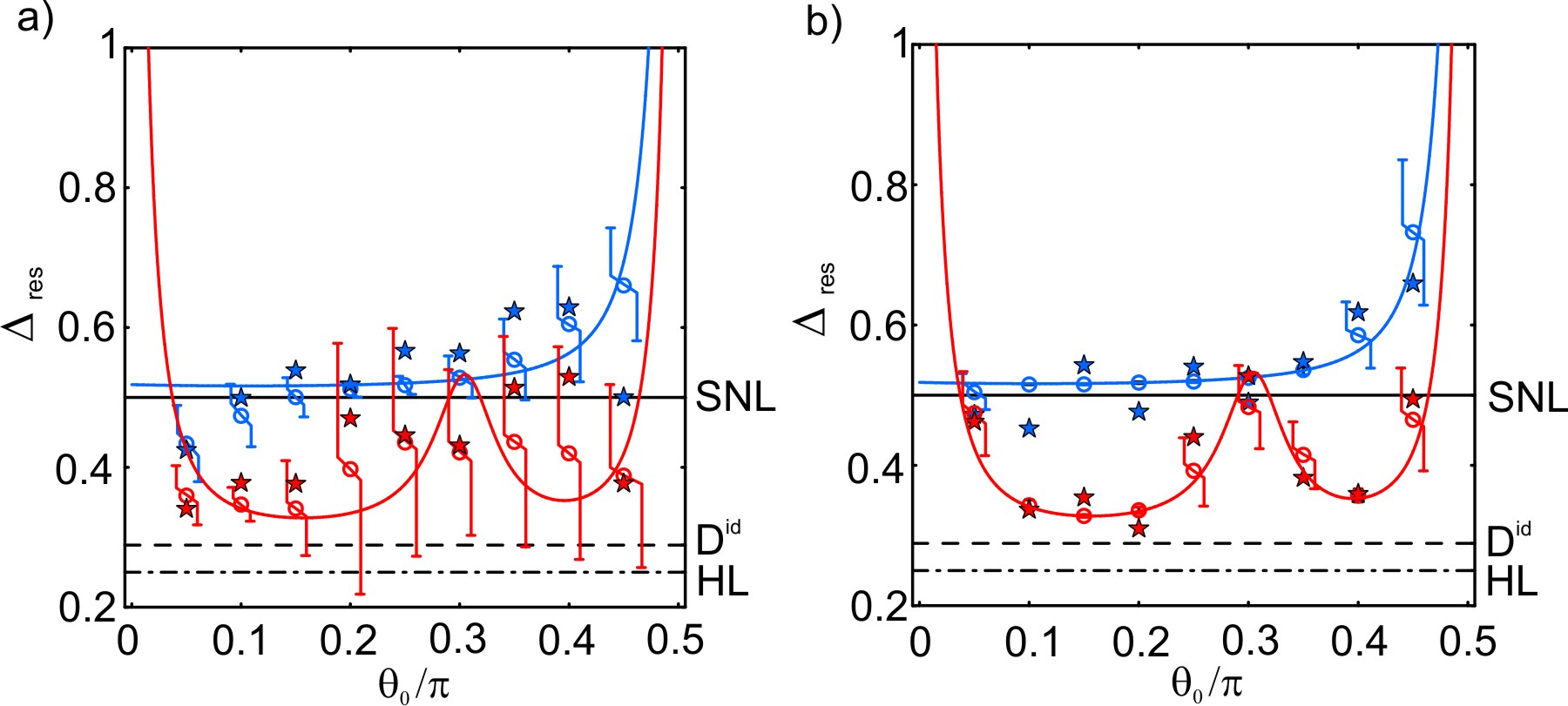}
\end{center}
\caption{
Rescaled phase uncertainties $\Delta_{\rm res}$ obtained with the 
probe state $\rho^{\rm exp}_D$ (red)
and $\rho^{\rm exp}_{\rm sep}$ (blue), with a) $m=10$ and b) $m=100$. 
The solid red (blue) line is the expected uncertainty  
given by the CRLB [Eq.~(\ref{eq:CR})] using
the experimental $F_{\hat\mu}$ [see Fig.~\ref{fig:Pfit} l)]
for the state $\rho^{\rm exp}_D$ ($\rho^{\rm exp}_{\rm sep}$).
Stars are the results of the ML analysis
(standard deviation of 
$\theta_{\mathrm{est}}$) with $\Delta_{\rm res}=\sqrt{m}\Delta\theta_{\mathrm{est}}$.
Circles are the results of the Bayesian analysis with $\Delta_{\rm res}=\sqrt{m}\mean{C}$, and the
error bars display the scatter of C.
Horizontal lines are the shot-noise limit (SNL) 
(solid line), 
the limit for an ideal Dicke state ($D^{\rm{id}}$) $\Delta_{\rm res}=1/\sqrt{F_Q[\ket{D_4^{(2)}}]}$ (dashed line) 
and the Heisenberg limit (HL) $\Delta_{\rm res}=1/4$ (dot-dashed line).}
\label{fig:Results}
\end{figure}

Fig.~\ref{fig:histograms} shows the distributions of the estimator 
$\theta_{\mathrm{est}}$ for the phase shift $\theta_0=0.2\pi$
and different values of $m$.
As expected, with increasing $m$,
the histograms approach a Gaussian shape with
standard deviation $\Delta\theta_{\mathrm{est}}$ decreasing as $1/\sqrt{mF_{\hat{\mu}} }$.
The width of the histograms is smaller for the Dicke state (red lines) 
than for the separable state (blue lines).
Fig.~\ref{fig:Results} shows $\Delta_{\rm res}=\sqrt{m}\Delta\theta_{\mathrm{est}}$ as a function of $\theta_0$.
For $m=10$ the standard deviation is below the CRLB~(Eq.~\ref{eq:CR}) for several $\theta_0$ values. 
This is possible because the estimation is biased, {\em i.e.},
for $b\equiv \mean{\theta_{\mathrm{est}}} - \theta_0$ we have $b \neq 0$ and 
$\partial_{\theta_0} b \neq 0$ \cite{Cramer_book,SI}.
The bias can be taken into account by replacing the numerator in the 
CRLB Eq.~(\ref{eq:CR}) by $|1-\partial_{\theta_0} b|$.
For even smaller $m$, only few different maxima of the likelihood functions 
${\cal L}(\theta)$ can occur, see Fig.~\ref{fig:histograms} a).
Then, $\theta_{\mathrm{est}}$ scatters significantly and 
hardly allows for an unbiased phase estimate.
When $m=100$, the bias
is strongly reduced
and the agreement of $\Delta \theta_{\mathrm{est}}$ with the unbiased CRLB is improved significantly.
While the bias is still large enough to cause apparent sensitivities 
below the shot-noise limit for the separable state, for the Dicke state the CRLB 
is saturated for a large phase interval.
This clearly proves that the multiparticle 
entangled Dicke state created experimentally
indeed achieves the SSN phase uncertainty predicted by the CRLB
Eq.~(\ref{eq:CR}) using the experimentally obtained Fisher information from Fig.~\ref{fig:Pfit} l).

A conceptually different phase estimation protocol is given by the
Bayesian approach assuming that the phase shift is a random variable.
The probability density for the true value of the phase shift being equal to $\theta$,
conditioned on the measured results $\mu_1,\mu_2,...,\mu_m$,
is provided by Bayes' theorem, $P(\theta|\{\mu_i\}_{i=1}^m)=P_{\rm exp}(\{\mu_i\}_{i=1}^m|\theta)P(\theta)/P(\{\mu_i\}_{i=1}^m)$. 
To define the {\em a priori} probability density $P(\theta)$ we adopt
the \emph{maximum ignorance principle} and take $P(\theta)$ to be
constant in the phase interval considered. 
The Bayesian probability density is then given by 
$P(\theta|\{\mu_i\}_{i=1}^m) \propto \prod_{i=1}^m P_{\rm exp}(\mu_i|\theta)={\cal L}(\theta)$. 
The phase shift can be estimated as the maximum 
of the probability density as before.
However, in contrast to the ML method, 
the Bayesian analysis allows to assign a meaningful 
uncertainty to this estimate even for a single $m$-experiment
and biased estimators.
This can be taken, for instance, as a confidence 
interval $\Delta\theta = C$ around the estimate, where the 
area of $P(\theta|\{\mu_i\}_{i=1}^m)$ is equal 
to $68\%$ (see Fig.~\ref{fig:histograms} d) and \cite{SI}).

Figs~\ref{fig:histograms} d)-f) illustrate how 
the Bayesian probability density evaluated for a single $m$-experiment
becomes Gaussian with a width $1/\sqrt{m F_{\hat\mu}}$, 
already for small values of $m$.
In contrast, the ML histograms [Figs~\ref{fig:histograms} a)-c)]
approach a Gaussian shape more slowly.
We also investigated how the Bayesian analysis performs on
average using the same data as in the ML case.
The results are shown in Fig.~\ref{fig:Results} with the rescaled Bayesian uncertainty $\Delta_{\rm res}=\sqrt{m}C$ 
for various $\theta_0$ and averaged over several $m$-experiments.
For $m=10$ the mean values of the 
confidences deviate from the CRLB and have 
a large spread. For $m=100$, however,
the confidences agree well with the CRLB
for most values of $\theta_0$, for both 
states.

In conclusion, we have investigated experimentally the relation between 
SSN phase estimation and the 
entanglement properties of a probe state.
We have identified useful 
multiparticle entanglement by determining
the quantum Fisher information from the
tomographical data of a four photon Dicke state. 
The benefit of such entanglement 
has been demonstrated by implementing 
two different phase estimation analyses, 
both of which saturate the Cramer Rao bound and 
clearly surpass the shot noise limit.
The approach is completely general: it
applies for any probe state, is scalable in the number 
of particles and does not require state selection. 
Our study thus provides a guideline for the future technological 
exploitation of multiparticle entanglement to 
outperform current metrological limits.

We thank N. Kiesel, W. Laskowski, and O. G{\"u}hne 
for stimulating discussions. R.K., C.S., W.W., and H.W. acknowledge the support
of the DFG-Cluster of Excellence MAP, the EU projects QAP and Q-Essence, and the 
DAAD/MNISW exchange program. W.W. and C.S. thank QCCC of the 
Elite Network of Bavaria and P.H. thanks the ERC Starting Grant GEDENTQOPT.



\begin{thebibliography}{99}
\expandafter\ifx\csname natexlab\endcsname\relax\def\natexlab#1{#1}\fi
\expandafter\ifx\csname bibnamefont\endcsname\relax
  \def\bibnamefont#1{#1}\fi
\expandafter\ifx\csname bibfnamefont\endcsname\relax
  \def\bibfnamefont#1{#1}\fi
\expandafter\ifx\csname citenamefont\endcsname\relax
  \def\citenamefont#1{#1}\fi
\expandafter\ifx\csname url\endcsname\relax
  \def\url#1{\texttt{#1}}\fi
\expandafter\ifx\csname urlprefix\endcsname\relax\def\urlprefix{URL }\fi
\providecommand{\bibinfo}[2]{#2}
\providecommand{\eprint}[2][]{\url{#2}}


\bibitem{GiovannettiNatPhot11}
For a recent review see
V. Giovannetti, S. Lloyd, and L. Maccone, Nat. Phot. {\bf 5}, 222 (2011).

\bibitem{RarityPRL90}
J.G. Rarity $et\,al.$,
Phys. Rev. Lett. {\bf 65}, 1348 (1990).

\bibitem{MitchellNat04} M.W. Mitchell, J.S. Lundeen and A.M. Steinberg,
Nature {\bf 429}, 161 (2004);
P. Walther $et\,al.$,
$ibid.$ {\bf 429}, 158 (2004)

\bibitem{KacprowiczNP2010} M. Kacprowicz $et\,al.$, 
Nature Phot. {\bf 4} 357 (2010).

\bibitem{NagataSci07} T. Nagata, $et\,al.$, 
Science {\bf 316}, 726 (2007).

\bibitem{Xiang_2010} Xiang, G.~Y., \emph{et al.,} 
{\it Nature Phot.} {\bf 5} 43 (2010).

\bibitem{AppelPNAS2009} J. Appel $et\,al.$, 
PNAS {\bf 106}, 10960 (2009);
M.~H. Schleier-Smith $et\,al.$,
Phys. Rev. Lett. {\bf 104} 073604 (2010).

\bibitem{LeibfriedSci04} D. Leibfried $et\,al.$,
Science {\bf 304}, 1476 (2004).
  
\bibitem{GrossNat10} C. Gross $et\,al.$,
Nature {\bf 464}, 1165 (2010).

\bibitem{RiedelNat10} M.~F. Riedel $et\,al.$,
Nature {\bf 464}, 1170 (2010). 

\bibitem{ReschPRL07}
K.J. Resch $et\,al.$,
Phys. Rev. Lett. {\bf 98}, 223601 (2007).


\bibitem{PezzePRL09}L. Pezz\'e and A. Smerzi, 
Phys. Rev. Lett. {\bf 102}, 100401 (2009).

\bibitem{HyllusArXiv10b}
P. Hyllus $et\,al.$,
http://arxiv.org/abs/1006.4366; 
G. T{\'o}th,
http://arxiv.org/abs/1006.4368.

\bibitem{Helstrom67} C.W. Helstrom, 
Phys. Lett. {\bf 25A}, 101 (1967);
S.L. Braunstein and C.M. Caves, 
Phys. Rev. Lett. {\bf 72} 3439 (1994).

\bibitem{SI}
See supplementary material for additional information.

\bibitem{nota_k-ent} A pure state is called $k$-particle entangled 
if $\ket{\psi^{k-{\rm ent}}}=\bigotimes_{l=1}^M \ket{\psi_l}$,
where $\ket{\psi_l}$ is a non-factorizable state of $N_l\le k$
particles, and $N_l=k$ for at least one $l$ \cite{SI}.

\bibitem{GHZ}
D.M. Greenberger,  M.A. Horne and A. Zeilinger, \emph{Going beyond Bell's theorem}
(Kluwer Academics, 1989).

\bibitem{GiovannettiPRL06} V. Giovannetti, S. Lloyd and L. Maccone,
Phys. Rev. Lett. {\bf 96}, 010401 (2006).
  
\bibitem{Cramer_book} H. Cram{\'e}r, \emph{Mathematical Methods of Statistics}
(Princeton Univ. Press, 1946).

\bibitem{note_resources}
Note that it is possible to further improve the 
sensitivity by applying the phase shift several
times to the probe system
\cite{GiovannettiPRL06}.


\bibitem{KieselPRL07}
N. Kiesel $et\,al.$,
Phys. Rev. Lett. {\bf 98}, 063604 (2007). 

\bibitem{notaFock} For indistinguishable qubits
the symmetric Dicke state reduces to a Twin-Fock state, see 
M.J. Holland and K. Burnett, 
Phys. Rev. Lett. {\bf 71} 1355 (1993).  

\bibitem{WieczorekPRL09}
W. Wieczorek $et\,al.$,
Phys. Rev. Lett. {\bf 103}, 020504 (2009).
  
\bibitem{HyllusPRA10} P. Hyllus, O. G{\"u}hne and A. Smerzi,
Phys. Rev. A {\bf 82}, 012337 (2010).

\bibitem{PezzePRL07}L. Pezz{\'e} $et\,al.$, 
Phys. Rev. Lett. {\bf 99}, 223602 (2007);
Z. Hradil $et\,al.$,
Phys. Rev. Lett. {\bf 76}, 4295 (1996).


  

        
\end{thebibliography}
\end{document}